\documentclass[12pt]{article}
\usepackage{graphics, color}
\usepackage{amssymb}

\def\XXint#1#2#3{{\setbox0=\hbox{$#1{#2#3}{\int}$} 
\vcenter{\hbox{$#2#3$}}\kern-.5\wd0}}

\begin{document}


\begin{titlepage}
\bigskip
\rightline{}
\bigskip\bigskip\bigskip
\centerline{\Large \bf Open Heterotic Strings}
\bigskip\bigskip\bigskip

\centerline{\large Joseph Polchinski}
\bigskip
\centerline{\em Kavli Institute for Theoretical Physics} 
\centerline{\em University of California} 
\centerline{\em Santa Barbara, CA 93106-4030} \centerline{\em joep@kitp.ucsb.edu}
\bigskip
\bigskip
\bigskip\bigskip


\begin{abstract}
We classify potential cosmic strings according to the topological charge measurable 
outside the string core.  We conjecture that in string theory it is this charge that governs the stability of long strings.  This would imply that the $SO(32)$ heterotic string can have endpoints, but not the $E_8 \times E_8$ heterotic string.  We give various arguments in support of this conclusion.
\end{abstract}
\end{titlepage}
\baselineskip = 17pt

\section{Stability of cosmic strings}

Compactifications of string theory give rise to many potential cosmic strings, including the fundamental strings themselves, D-strings and wrapped D-branes, solitonic strings and branes in ten dimensions, and magnetic and electric flux tubes in the four-dimensional effective theory.  The small value of the cosmological constant suggests that the compactified dimensions in our vacuum have great topological complexity~\cite{Sak}; consequently there may be of order $10^3$ cosmic string candidates, plus the bound states of these!  However, the only strings that are cosmologically relevant are those those actually produced, and moreover it is necessary that some take the form of infinite random walks in order to seed the later string network~\cite{Kib}.  Thus it is the much smaller number of phase breaking transitions after inflation that is relevant.  

Even if some infinite strings are produced, there is the possibility that they will decay too soon to be observed~\cite{Witcos,PV}.  In this paper we consider the stability of long strings.  We propose a topological classification, based on the topological charge observable {\it outside} the string core, and we conjecture that all decays allowed by this classification actually occur in string theory.  The rate might be slow on cosmological time scales; we are concerned here only with absolute stability.

We distinguish four kinds of macroscopic string, in three noncompact spatial dimensions:
\begin{itemize} 
\item
{\it Local} strings have no topological charge that can be detected by any measurement outside the core of the string.
 \item
{\it Global} strings have a topological charge that is visible in local measurements, the gradient of a scalar field that winds around the string.
\item
{\it Aharonov-Bohm} strings~\cite{AWil} have no topological charge that can be detected by local measurements, but have an Aharonov-Bohm phase with respect to at least one particle that is neutral under all massless low energy gauge groups.
\item
{\it Quasi-Aharonov-Bohm} strings have an Aharonov-Bohm phase, but only with respect to particles that are charged under the low energy gauge group. 
\end{itemize}
Note that this classification is along a different axis from the list in the first paragraph above, and applies to all strings in that list; it is independent of their internal structure.  In particular the earlier description changes under duality, and in the middle of moduli space the distinctions disappear, while the topological classification is duality-invariant.

The topological classification, which is a refinement of earlier discussions, determines the potential instabilities of the string:
\begin{itemize}
\item
For local strings there is no obstacle to the string breaking at many points and disappearing into a cloud of short {\it open} strings.  
\item
For global strings causality prevents this decay, since the topological charge can be seen at arbitrary distance, but these strings suffer another instability.   The nonexistence of continuous global symmetries in string theory~\cite{BD} implies that the scalar field will acquire a field-dependent potential, so that the string has a potential energy cost at long distance.  This ultimately makes it unstable: finite loops collapse, and even infinite strings disappear by annihilation when the typical transverse separation drops below the horizon scale~\cite{DW}. 
\item
{Aharonov-Bohm} strings are stable: since the phase can be measured at arbitrary distance, they cannot just disappear.
\item
For quasi-Aharonov-Bohm strings, there is a flux tube in the low energy gauge theory having exactly the same Aharonov-Bohm phases.  The quasi-Aharonov-Bohm string can decay to such a flux tube (by a local process, as we will see), and the tube then spreads radially outward and becomes a diffuse field.
\end{itemize}

We conjecture that in string theory every decay allowed by this classification actually occurs with nonzero rate.  One might have thought that the internal structure of the string core would play some role, but we believe that it does not.  We first illustrate the classification with some examples from field theory, and then discuss a few caveats.
\begin{enumerate}
\item
Consider first the Abelian Higgs model: a $U(1)$ gauge field and a Higgs field of charge~1.  According to the classification, the Abrikosov-Nielsen-Oleson vortices of this theory are local strings:  the only available probe particle, the Higgs field itself, is single-valued around the vortex.  These strings are absolutely stable, showing that the conjecture does not hold in arbitrary field theories.  
\item
Of course, when we see a $U(1)$ gauge symmetry at some scale we expect it to be embedded in a larger symmetry, and then there will be magnetic monopoles.  This is true even in less complete
unifications such as GUTS and Kaluza-Klein theory.  If we add a monopole of the minimum Dirac quantum, its flux $\int F = 2\pi$ is the same as carried by the string core.  A string can now end on a monopole, and a long string will decay by monopole-antimonopole production.  The string is still local, because the monopole feels no phase from the flux in the core.

This illustrates the point of classifying defects according to the externally measurable charge: it is only this that is stable against embedding into a more complete theory.
\item
If instead we add only monopoles whose flux is an integer multiple $q$ of the minimum, then 
single vortices can no longer end on the monopole, and are stable.  Rather, the monopole is a vertex where $q$ strings can end.
\item
If we again saturate the Dirac quantization by adding additional particles of charge $1/q$ to the previous case, we have an Aharonov-Bohm string, because these particles pick up a phase $2\pi/q$ in going around the string.  This string is still stable.
\item
Now consider the Abelian Higgs models with an additional unbroken $U(1)'$ gauge field, and in addition to the Higgs of charge $(1,0)$ introduce particles of charge $(\frac{1}{2},\frac{1}{2})$.  Monopoles of magnetic charge $(2,0)$ and $(1,1)$, and integer linear combinations of these, will saturate the Dirac quantization condition.  The string, with magnetic fluxes $(2\pi,0)$, is now quasi-Aharonov-Bohm, since it can be seen by the $(\frac{1}{2},\frac{1}{2})$ particles but not by the $U(1)'$-neutral $(1,0)$ particles.  

The string can break into open strings by pair production of $(1,1)$ monopoles.  These have long-range $U(1)'$ flux, and flux lines run from the right endpoint of each open string to the left endpoint of the next.  When the open strings shrink to zero length, these flux lines join into a $U(1)'$ magnetic flux tube which can then spread freely.
\end{enumerate}

We see that when the Dirac quantization is saturated (cases 2, 4, and 5), the topological classification indeed determines the stability of the string.  In string theory it is strongly suspected, though not proven, that the quantization condition is saturated~\cite{Joemon}.\footnote{Additional conjectures about the spectrum of charges appear in ref.~\cite{AMNV}.}  It would be interesting to find a perspective from which the result is obvious.  Proving it on a case-by-case basis is challenging; for example, in ref.~\cite{WW} determining the quantization of monopole charge in Calabi-Yau compactifications required an intricate topological analysis.  In ref.~\cite{JoeRR} it followed from a seeming unrelated string loop calculation.  In the next two sections we will examine other cases where the conjecture is not obvious but turns out to be true.

Now we consider the caveats.  Another possible decay is that of semilocal strings, reviewed in ref.~\cite{AV}.  This is similar to that of quasi-Aharonov-Bohm strings, except that it is a scalar field configuration that spreads.  However, this depends on having an exact global symmetry, and so cannot occur in string theory, though in some cases a string might be able to lower its energy by a limited amount of spreading.  Note that radial spreading is consistent with causality; it is dynamical considerations that determine whether it can occur.  There is also the semilocal case where all symmetries are gauged; we believe that this will fall under our classification.

We have not provided a precise definition of topological charge.  For example, the deficit angle of any string implies a spacetime holonomy.\footnote{This was pointed out by B. Pioline.}  We exclude this from the definition because it is also produced by a line of point particles.  A string can also have a dilaton charge, but the dilaton does not wind so this is not topological.

Global strings will be stable in some supersymmetric compactifications, because supersymmetry can protect flat directions in the potential.  They will be unstable in any nonsupersymmetric vacuum.  However, as far as is now known all such vacua are themselves unstable, so the absolute stability of strings is no longer meaningful!  The classification still retains practical significance: once one identifies a vacuum that is stable on cosmological time scales, one must examine the allowed decays of any potential cosmic string.

\section{The $SO(32)$ heterotic string}

In ref.~\cite{Witcos} it was noted that heterotic strings in the simplest compactifications will be global strings, because of their coupling to the Kalb-Ramond field $B_{\mu\nu}$ which in four dimensions is dual to a scalar.  However, ref.~\cite{WitO32} identifies a mechanism, anomaly cancellation, that removes the $B_{\mu\nu}$ field from the spectrum in models whose fermion content has a $U(1)$ gauge anomaly.  The effect of such pseudoanomalous $U(1)$'s on macroscopic heterotic strings was considered in ref.~\cite{HarvNac}.  We will review their results in more detail below, but for now we note one conclusion: one can dress the heterotic string in such a way that it has no long-range local fields.  Thus it must be either local, Aharonov-Bohm, or quasi-Aharonov-Bohm, and we will see that all three types are possible depending on the model.
This raises a paradox.  It is well-known that there are no open heterotic strings~\cite{PSQ}, because there is no consistent boundary condition relating the left- and right-movers.
Thus we appear to have an exception to the stability conjecture.  In fact, we will argue that it is the integrity of the heterotic string that must yield.  In this section we consider the $SO(32)$ heterotic string, and in the next the $E_8 \times E_8$ heterotic string.

\subsection{Ten dimensions}

The discussion so far has dealt with compactified string theories, but if open heterotic strings exist in these then they must exist in ten dimensions also, by consideration of the large-radius limit.  Therefore we will discuss first the noncompact theory, and then return to the compactifications discussed above.

The first obstacle to the breaking of a heterotic string is Kalb-Ramond charge conservation. The integral of the field strength $*H$ on a seven-sphere surrounding the heterotic string is nonzero, and this is inconsistent if the seven-sphere can be pulled off the end of an open heterotic string.
Equivalently, the action is not invariant under the Kalb-Ramond gauge transformation $\delta B = d\alpha$ if the world-sheet has a boundary.  The relevant term in the world-sheet action transforms as 
\begin{equation}
\delta \frac{1}{2\pi\alpha'} \int_{W} B =  \frac{1}{2\pi\alpha'} \int_{\partial W} \alpha\ ,
\end{equation}
where $W$ is the string world-sheet.

The remedy here is known from many examples~\cite{CS,Witbound, Stromopen}.  The field strength $*H$ must also be sourced by fields in the spacetime action, in such a way that the total source is conserved and the total action is gauge invariant.  The relevant term in the action is
\begin{equation}
\frac{1}{4! (2\pi)^5 \alpha'} \int B\, {\rm Tr}_v F^4\ ,
\end{equation}
as required by anomaly cancellation~\cite{GS}.  The $SO(32)$ trace is in the vector representation.  There are also curvature terms, which would be relevant for a string ending at a conic singularity, but not for one ending at an ordinary point of spacetime.  Thus the invariance is restored if $d\, {\rm Tr}_v F^4$ is nonzero on the string boundary, or equivalently if
\begin{equation}
\frac{1}{4! (2\pi)^4} \int_{S^8}  {\rm Tr}_v F^4 = \pm 1 \label{flux}
\end{equation}
for any eight-sphere enclosing a single heterotic string endpoint; the sign must be opposite for the two ends of the open heterotic string.  This is similar to the way that $B$-gauge invariance is preserved for strings ending on a D-brane, though in that case $\int_{S^8} *F$ is the relevant invariant.  Gauge fields with nonzero $\int_{S^8}  {\rm Tr}\, F^4$ have been discussed in refs.~\cite{CDFN,GKS,Tchr,FN1,FN2}; various applications of these in string theory are given in refs.~\cite{DuffLu,HarvStr,DDP,DKL,Horava,OSz,MSV,BHTZ}.

The foregoing is similar for any open string or brane, but for an open heterotic string there is another apparent obstacle.  There are more left-moving than right-moving degrees of freedom on the string, and so there is no sensible boundary condition that can be imposed: degrees of freedom are continually gained at the right endpoint and lost at the left endpoint.  

This problem is remedied in an interesting way.   Consider a semi-infinite string, whose single endpoint is placed at the origin; we may assume that the vector potential is transverse, $\vec r \cdot \vec A = 0$.  The Dirac equation for a chiral spinor field in spacetime, transforming under some representation $R$ of the gauge group, is
\begin{equation}
\left(\gamma^t  \partial_t + \gamma^r \partial_r + D \hspace{-8pt} / \hspace{2pt}^R \right) r^{4} \Psi = 0\ .
\end{equation}
Here $D \hspace{-8pt} / \hspace{2pt}^R$ is the Dirac operator along the $S^8$ surrounding the endpoint.
The product $\gamma^t \gamma^r$ is equal to the chirality in the $S^8$ subspace.
The zero modes of $D \hspace{-8pt} / \hspace{2pt}^R$ move radially outward or inward, depending on the sign of $\gamma^t \gamma^r$.  Thus the index, given by the Atiyah-Singer index theorem as
\begin{equation}
{\rm Tr}\, \gamma^t \gamma^r = \frac{1}{4! (2\pi)^4} \int_{S^8}  {\rm Tr}_R F^4\ ,
\end{equation}
measures the mismatch between the outgoing and ingoing modes.  If $R$ were the vector representation this mismatch would then be $1$ (we will not attempt to keep track of the sign, but assume that it works out correctly).  In the heterotic string $R$ is the adjoint representation of $SO(32)$.  The mismatch is then 24, from ${\rm Tr}_{adj} F^4 = (32 - 8) {\rm Tr}_v F^4 +$ double trace terms; the double trace terms make no contribution on $S^8$.

A similar mismatch occurs for magnetic monopoles in the effective field theory of the Standard Model, leading to the Rubakov-Callan effect~\cite{Rub,Cal}.  In that case, the numbers of ingoing and outgoing modes are the same but their $SU(3) \times SU(2) \times U(1)$ quantum numbers differ.  Preservation of the gauge symmetries forces the boundary condition at the origin to violate baryon number, allowing the monopole to catalyze baryon decay with a large cross section.

Here even the numbers mismatch, and there is no consistent boundary condition for the spacetime fermions at either endpoint.  This is of course the same problem that we had on the world-sheet, and there the mismatch (in a physical gauge) was 32 left-moving current algebra fermions minus 8 right-moving Ramond-Neveu-Schwarz or Green-Schwarz fermions, a net of 24.  Thus there is an obvious resolution: there are 24 modes that flow from spacetime into the right endpoint, move to the left on the world-sheet, and reenter spacetime at the left endpoint!  

For example, we can take a gauge field that lies in an $SO(8)$ subgroup and leaves unbroken
an $SO(24)$ gauge symmetry and an $SO(8)$ diagonal sum of spacetime and gauge rotations.
The spacetime zero modes transform as a ${\bf 24}$ of the gauge group, $\Lambda^a$, and so we must have at the endpoints $\Lambda^a = \pm \lambda^a$, where the $\lambda^a$ are the current algebra fermions for the unbroken group.  Note that the GSO projection on $\lambda$ must be retained.  The remaining eight $\lambda$ do reflect off the endpoint and become the eight supersymmetric fermions; the boundary condition is determined by conservation of the world-sheet current for the diagonal symmetry.

Now let us consider energetics.  The condition~(\ref{flux}) implies that the field strength falls off as $r^{-2}$, and so the energy density falls as $r^{-4}$ but the total energy diverges as $r^4\, dr$.\footnote{We are assuming that the gauge field is spherically symmetric.  It could also take the form of a thick string, such as that in ref.~\cite{HarvStr} (we thank A. Strominger for this comment).  Likely the spherically symmetric configuration is unstable to collapse toward a less symmetric one.}
  For a pair of endpoints separated by a distance $l$ one therefore expects a potential of order $l^5$.
Thus, in the terminology of ref.~\cite{CMP}, the heterotic string in ten dimensions can {\it end} but not {\it break}.  That is, there are configurations such that the endpoint is topologically allowed, but the spontaneous formation of a pair of endpoints in a long string is energetically forbidden.  The situation will be different in compactified theories.

We see that there is no objection in principle to the heterotic string having endpoints.  It should be emphasized that the open heterotic string cannot be treated via ordinary string perturbation theory, because the boundary conditions mix the first-quantized world-sheet fields with the second-quantized spacetime fields; this is intrinsically nonperturbative.  Note also that the $B$ gauge invariance involves cancellation between the tree level world-sheet action and a one-loop term in spacetime.  The absence of a perturbative description means that the open heterotic strings are not expected to lead to new massless states.  

We have avoided the right-left mismatch of world-sheet gauge symmetries by considering a physical gauge.  In a covariant gauge we would again have a mismatch, with more ghost and matter fields moving to the right: we cannot lift the endpoint to the larger Hilbert space in which the constraints act.  However, this is also true of weak-strong dualities --- the enlarged Hilbert space is a technical artifact and dualities do not act on it.

Thus far we have motivated the existence of the open heterotic string, but not given an explicit construction.  We can do better in the $S$-dual description of the $SO(32)$ theory, where the open heterotic string becomes a type~I D1-brane~\cite{PW} ending on a type~I D9-brane.  We can describe this along the lines of ref.~\cite{SenK,WitK}, working in a sector of 16 D9-branes and 8 anti-D9-branes.  The gauge group is $SO(16) \times SO(8)$ and the tachyon $T$ (which is real in the type I theory) transforms as $({\bf 16},{\bf 8})$.  The tachyon vacuum manifold is 
\begin{equation}
T_0 = \left[\begin{array}{c} I_{8\times 8} \\ 0_{8\times 8} \end{array} \right]
\end{equation}
and its $SO(16)$ rotations.  Each vacuum can thus be regarded as a map from ${\bf R}^8$ into ${\bf R}^{16}$, which reduces to a map from $S^7$ into $S^{15}$.  

Now consider the vacuum as a function $T(\Omega)$ on the $S^8$ surrounding the endpoint, independent of the radius $r$.  We take a configuration in which this is equal to the fourth Hopf map, which represents $S^{15}$ as an $S^7$ bundle over $S^8$ (see e.g. ref.~\cite{Baez}).  This bundle is nontrivial, and so as a function $T(\Omega)$ must be singular as we approach one pole of $S^8$, which we choose to lie on the positive $x^9$ axis.  
If we take a basis for the $SO(9)$ $\gamma$-matrices such that 
\begin{equation}
\gamma^9 = \left[\begin{array}{cc} I & 0 \\ 0 & -I \end{array} \right], \quad
\gamma^i = \left[\begin{array}{cc} 0 & \Gamma^i \\ (\Gamma^i)^T & 0\end{array} \right]\ ,\ i = 1,\ldots, 8 \ ,
\end{equation}
then an explicit form for the tachyon field is
\begin{equation}
T(\Omega) = \frac{ r - \sum_{i=1}^9 \gamma^i x^i }{\sqrt{2r(r-x^9)} } T_0\ .
\end{equation}
Now we smooth $T(\Omega)$ by multiplying by $f(l)$, where 
$f(0) = 0$ and $f(l) $ approaches $1$ rapidly at large $l$.  Here 
$l$ is taken to be the distance to the nearest point on the positive $x^9$ axis (this also smoothes the field near the origin).   Thus, $T(\Omega)$ is a well-defined function in spacetime, not just a section.  We now have a smooth tachyon field which is a vacuum configuration everywhere except near the positive $x^9$ axis.  The field near the axis is
\begin{equation}
T \approx f(l) \left[\begin{array}{c} 0 \\  \sum_{i=1}^8 \Gamma^i x^i/l \end{array} \right]\ .
\end{equation}
This is just the D1-brane as constructed in ref.~\cite{WitK}, and so this configuration is semi-infinite type~I D1-brane.  Open string field theory therefore provides a construction of this object on the type~I side, along the lines of refs.~\cite{HarK,BSZ,Jev,SFTSen,SFTWit}.  Alternately, it seems likely that this construction can be embedded in heterotic matrix theory (see ref.~\cite{Banks} for a review and references).

The $SO(16)$ gauge field satisfies $(\partial_i - i A_i) T = 0$ so as to cancel the $r^{-2}$ part of the energy density, except that like $T$ it is smoothed near the origin and the positive $x^9$ axis.  This undetermines $A_i$: we can vary the connection in the unbroken $SO(8)$; note that the field strength is always in this $SO(8)$.  That $\int_{S^8}  {\rm Tr} F^4$ takes the correct value can be deduced as follows.  For the smoothed configuration this integral must vanish, because the ${S^8}$ can be shrunk smoothly to a point.  The integral omitting the D-string core near the positive $x^9$ axis is then equal to minus the integral in the core, which is correctly normalized~\cite{WitK}.

From the preceding discussion it follows that we should perhaps think of the heterotic theory as containing NS9-branes, as suggested in ref.~\cite{Hull} for other reasons.  Note that the energy density of the gauge fields at the endpoint is of order $g^{-2}$, so the endpoints are more like ordinary solitons than D-objects.  Conceivably the quantization of short open heterotic strings could lead to long-lived states with an energy parametrically between the $g^{0}$ of the perturbative states and the $g^{-2}$ of the classical endpoints, associated with the expected $e^{-1/g}$ effects~\cite{Shenker,Silverstein}.

\subsection{Four dimensions}

Now let us consider the compact theory.  A simple example of a compactification with an anomalous $U(1)$ is the $SO(32)$ heterotic string on a Calabi-Yau manifold with spin connection equal to gauge connection~\cite{WitO32,HarvNac}.  The spectrum under the $SO(26) \times U(1)$ that commutes with the $SU(3)$ holonomy is $b_{1,1}$ copies of ${\bf 26}_1 \oplus {\bf 1}_{-2}$ and $b_{2,1}$ copies of ${\bf 26}_{-1} \oplus {\bf 1}_{2}$.  Noting that
\begin{equation}
b_{2,1} - b_{1,1} = \frac{\chi}{2} =
\frac{1}{3! (2\pi)^3 } \int {\rm Tr}_v \tilde F^3\ ,
\end{equation}
where $\tilde F$ refers to the background field, we see that the ten-dimensional anomaly cancellation term descends to the four-dimensional term
\begin{equation}
\frac{\chi}{(2\pi)^2 \alpha'} \int BF \label{bf}
\end{equation}
where $F$ is the $U(1)$ field strength.  This cancels the anomaly from the chiral fermions, and also breaks the $U(1)$ symmetry.

Now consider in the low energy field theory a magnetic monopole with
\begin{equation}
\int_{S^2} F = \pi \label{fluxpi}
\end{equation}
in the unbroken $U(1)$ and also in each factor in $U(1)^{13} \subset SO(26)$.
This is correctly quantized with respect to all fields (including the $SO(32)$ spinors), and is the minimum charge under the unbroken $U(1)$.  At the monopole center, where $dF$ is nonzero, the gauge invariance $\delta B = d\alpha$ breaks down.  This is exactly cancelled if $\frac{1}{2} \chi$ heterotic strings, appropriately oriented, end at the monopole: evidently the number of heterotic strings is conserved modulo $\frac{1}{2} \chi$.  In particular, a single heterotic string can break by formation of a monopole-antimonopole pair precisely if $|\chi| = 2$.  For larger values of $\chi$, which must be even, the monopole is a multi-string junction.

The monopole has a finite energy: the integral of the flux energy converges both at long distance and at short distance (where it goes over to the ten-dimensional calculation).  The integral is dominated by the compactification scale, and so is of order $R^5 /\alpha'^3 g_{\rm s}^2$.  Suppression of the decay on cosmological time scales only requires that this be an order of magnitude above the square root of the string tension, which can easily be the case.

We can again see that the Rubakov-Callan effect operates to produce consistent boundary conditions.  The four-dimensional index is
\begin{equation}
{\rm Tr}\, \gamma^t \gamma^r = \frac{1}{2\pi} \int_{S^8}  {\rm Tr}_R F = 6 {\chi}\ .
\end{equation}
We are now dealing with Weyl fields, so for $\chi = 2$ the mismatch is 24 Majorana outmovers vs. inmovers, allowing consistent boundary conditions when a single heterotic string ends.  For larger values of $\chi$, the spacetime mismatch is precisely equal to the total mismatch on the attached strings at the junction, allowing consistent boundary conditions.  Indeed there are many choices of consistent boundary condition; understanding the detailed dynamics of the junction is a challenging problem.

To see where these strings fit into our earlier taxonomy, let us examine the fields outside a long string.  In theories with anomalous $U(1)$s, low energy supersymmetry implies a $D$-term which generally leads to further gauge symmetry breaking.  However, it is simplest to imagine first the case that supersymmetry breaking stabilizes the point where all other charged fields vanish.
A straight string in the 1-direction couples to the potential $B_{01}$, which couples in turn to the field strength $F_{23}$.  The field equations, assuming for simplicity a flat metric and constant dilaton, are
\begin{eqnarray}
\frac{1}{2\kappa^2} \partial_\perp^2 B_{01} &=& \frac{1}{2\pi\alpha'} \delta^2(x^\perp)
+ \frac{\chi}{(2\pi)^2 \alpha'} F_{23}\ , \label{eom} \\
\frac{1}{g^2} F_{23} &=& \frac{\chi}{(2\pi)^2 \alpha'} B_{01}\ .
\end{eqnarray}
The second line is the field equation from $A_\mu$, with the requirement that the fields vanish at infinity.  The fields fall expontially for nonzero $\chi$, with a mass-squared of order $\kappa^2 g^2 /\alpha'^2 \sim g_{\rm s}^4/\alpha'$.  The string is thus shielded by a surrounding flux tube, whose radius is much larger than the string scale for small $g_{\rm s}$.  Integrating eq.~(\ref{eom}) over the transverse plane, the left-hand side must vanish and so the total flux in the tube is
\begin{equation}
\int_{R^2_\perp} F = \frac{2\pi}{\chi}\ . \label{transflux}
\end{equation}

Now consider the unstable flux tube, $|\chi| = 2$, giving total flux $\pm \pi$.  There is no local long-range field, but the flux can be detected with a probe particle of unit charge, for example the ${\bf 26}_{\pm 1}$.  However, these are charged under the unbroken $SO(26)$.  The ${\bf 1}_{\pm 2}$ fields do not couple to the unbroken group but also do not feel the flux.  This string therefore quasi-Aharonov-Bohm, consistent with our conclusion that it is unstable.  What happens in detail is that the short open strings that form have a $U(1)$ gauge field that runs coaxially, but also a long-range $SO(26)$ magnetic field running from the right endpoint of each string to the left endpoint of the next.  Eventually the monopole-antimonopole at the endpoints of each open string annihilate, and one is left with a dissipating $SO(26)$ flux tube.

For larger values of $\frac{1}{2} |\chi|$, the ${\bf 1}_{\pm 2}$ can detect the flux and so we have an Aharonov-Bohm string, consistent with the absolute stability.  Note that the $U(1)$ flux of the monopole runs out coaxially with the $\frac{1}{2} |\chi|$ attached strings, while the $SO(26)$ flux is a Coulomb field centered on the junction.

Including now the effect of the D-term potential, a ${\bf 1}_{\pm 2}$ scalar $Y$ will have a vacuum expectation value.  We could consider a heterotic string with topologically trivial $Y$, or a string built as a soliton in $Y$ with no heterotic core, or a bound state of these.  (These possibilities were discussed in ref.~\cite{HarvNac}, while refs.~\cite{CMMQ,Quiros,BDP} discussed the $Y$-strings).
Thus let us take the general case of a core consisting of $m$ heterotic strings, coaxial with a field configuration having winding $Y \propto e^{in\theta}$.  The field equation for $B_{01}$ is unchanged so it follows as above that the integrated flux is $2\pi m/\chi$, and $A_\theta = m/\chi$ at long distance.  The fields at infinity are therefore pure gauge if $n = 2m/\chi$.  If this condition is not satisfied then the string is global, and will be confined by instantons.  If it is satisfied, then the interaction of a ${\bf 1}_{\pm 2}$ probe is as before: there is a Bohm-Aharonov phase for $|\chi| > 2$, and the string is stable.  For $|\chi| = 2$ there is no Bohm-Aharonov phase for a neutral particle and the string should be unstable.  Taking e.g. $\chi = 2$, $m=n=1$, which is a $Y$-string coaxial with a heterotic string, the monopole~(\ref{fluxpi}) at the end of the heterotic string has unit flux with respect to $Y$, and so unwinds the $n=1$ $Y$ field and terminates the $Y$-string along with the heterotic string.

\section{The $E_8 \times E_8$ heterotic string}

The behavior of the $E_8 \times E_8$ heterotic string is different.  The $E_8$ algebra has no independent quartic Casimir~\cite{GS}, so ${\rm Tr}\, F^4$ is proportional to $({\rm Tr}\, F^2)^2$.  The latter vanishes when integrated on $S^8$, because $d{\rm Tr}\, F^2  = 0$ implies that ${\rm Tr}\, F^2  = d\omega$ for some globally defined form $\omega$ on $S^8$.  Thus the $E_8 \times E_8$ string {\it cannot} have endpoints in ten dimensions.\footnote{It might be that it could end at a singular point, such that the angular directions around the endpoint have the appropriate topology.}

This is perhaps surprising, but it not inconsistent with the $SO(32)$ result.  One might think that the $T$-duality of the two heterotic theories~\cite{narain} would lead to a contradiction, by starting with a long string in the toroidally compactified $SO(32)$ theory.  However, the $SO(32)$ string is global in this compactification, so we cannot in this way construct a broken $E_8 \times E_8$ string.

Now let us consider pseudoanomalous compactifications.  Following the discussion in ref.~\cite{WitO32}, the anomaly in a $U(1)$ generator $T$ is a linear combination of
\begin{equation}
\int {\rm Tr}\,T^3 \tilde F^3\ ,\quad \int  {\rm Tr}\, (T^3 \tilde F )({\rm Tr}\, R^2)\ .
\end{equation}
In $E_8$ these can be written in terms of the quadratic Casimir as a linear combination of 
\begin{equation}
\int ({\rm Tr}\, T \tilde F)^3\ ,\quad \int ({\rm Tr}\, T^2)(  {\rm Tr}\, T \tilde F )( {\rm Tr}\, \tilde F^2)\ , \quad
\int ({\rm Tr}\, T^2 )( {\rm Tr}\, T \tilde F )( {\rm Tr}\, R^2) \ .
\end{equation}
All of these involve ${\rm Tr}\, T \tilde F$, so the anomaly is nonzero only if the background field strength in the anomalous $U(1)$ is nonvanishing.  
The linearized Kalb-Ramond field strength is
\begin{equation}
H = dB - \frac{1}{30} A\, {\rm Tr}\, T \tilde F
\end{equation}
where $A$ is the anomalous $U(1)$ gauge field.  It follows that the component of $B$ proportional to ${\rm Tr}\, T \tilde F$ transforms nonlinearly under $U(1)$ gauge transformations.  Thus $A$ is Higgsed by both the axion dual to the spacetime components of $B$ and by one scalar coming from the internal components of $B$.
One linear combination of these scalars remains massless, and so the $E_8 \times E_8$ string is a global string when there are pseudoanomalous $U(1)$s.  This is consistent with the conclusion that it is stable against breakage.\footnote{More generally we could consider strings with independent winding numbers for the spacetime axion and the internal axion.  The cores of these would involve the moduli of the ten-dimensional compactification, so these would not be local objects in ten dimensions.  The analysis of these is beyond our present scope, but in some cases the string would not have a long-range field.}  The $SO(32)$ heterotic string would also be a global in compactifications with nonzero ${\rm Tr}\, T \tilde F$.

The $S$-dual picture would have been problematic if the $E_8 \times E_8$ string had an endpoint.  The heterotic string becomes a membrane extended in the eleventh dimension, whose endpoint would become an edge stretched between the boundaries of the M theory direction.  The gauge fields live only on the boundaries, so there would be no way to restore the $A_{\mu\nu\lambda}$ gauge invariance at the boundary of the membrane.  

\section{Discussion}

The somewhat surprising conclusions for the $SO(32)$ and $E_8 \times E_8$ strings give strong evidence that the stability conjecture is true.  It should be a useful guide in analyzing the stability of potential cosmic strings.  

We have seen that there are many compactifications in which the heterotic string is absolutely stable, and others in which it is very long-lived.  However, the problem remains that the tension is too large~\cite{Witcos,Turok}.  This is true even in the warped compactifications of refs.~\cite{warped,dWSHD}.  These are flat in the string metric, due to the property of general unitary CFTs that the spacetime translation currents are free world-sheet fields.  The fundamental heterotic string then does not feel the warping.\footnote{Other cosmic string candidates in heterotic string theory have recently been discussed in refs.~\cite{Buch,BB}.}

Finally, aside from the cosmological implications, these results may give new insight into the nonperturbative nature of heterotic string theory.

\subsection*{Acknowledgments}

I would like to thank N. Arkani-Hamed, M. Becker, E. Buchbinder, G. Dvali, A. Grassi, B. Pioline, E. Silverstein, and H. Tye for helpful conversations, and E. Silverstein and A. Strominger for detailed comments on the manuscript.  This work was supported by National Science Foundation grants PHY99-07949 and PHY00-98395.

\end{document}